\RequirePackage{lineno}
\documentclass[aps, prl,
floatfix,twocolumn,superscriptaddress,
]{revtex4-2}
\usepackage{t1enc}
\usepackage[T1]{fontenc}
\usepackage{lmodern}
\usepackage{graphicx}
\usepackage{epsfig}
\usepackage{amssymb}
\usepackage{amsmath}
\usepackage{dcolumn}
\usepackage{bm}
\usepackage{color}
\usepackage{float}
\usepackage{hyperref}
\usepackage{natbib}
\usepackage{footnote}
\usepackage{ulem}
\usepackage{hhline}
\begin{document}

\title{Linkage between scattering rates and superconductivity in  doped ferropnictides}

\author{J.\,Fink}
\affiliation{Leibniz Institute for Solid State and Materials Research  Dresden, Helmholtzstr. 20, D-01069 Dresden, Germany}
\affiliation {Max Planck Institute for Chemical Physics of Solids, D-01187 Dresden, Germany}
\affiliation {Institut f\"ur Festk\"orperphysik,  Technische Universität Dresden, D-01062 Dresden, Germany}
\author{ E.D.L.\,Rienks}
\affiliation{ Helmholtz-Zentrum Berlin, Albert-Einstein-Strasse 15, 12489 Berlin, Germany}
\author{M.\,Yao}
\affiliation {Max Planck Institute for Chemical Physics of Solids, D-01187 Dresden, Germany}
\author{R.\,Kurleto}
\affiliation {Leibniz Institute for Solid State and Materials Research  Dresden, Helmholtzstr. 20, D-01069 Dresden, Germany}
\affiliation {M. Smoluchowski Institute of Physics, Jagiellonian University, {\L}ojasiewicza 11, 30-348, Krak{\'o}w, Poland }
\author{J.\,Bannies}
\altaffiliation{Present address: Quantum Matter Institute, University of British Columbia, Vancouver, BC V6T 1Z4, Canada}
\affiliation {Max Planck Institute for Chemical Physics of Solids, D-01187 Dresden, Germany}
\author{S.\,Aswartham}
\affiliation{Leibniz Institute for Solid State and Materials Research  Dresden, Helmholtzstr. 20, D-01069 Dresden, Germany}
\author{I.\,Morozov}
\affiliation{Leibniz Institute for Solid State and Materials Research  Dresden, Helmholtzstr. 20, D-01069 Dresden, Germany}
\author{ S.\,Wurmehl}
\affiliation{Leibniz Institute for Solid State and Materials Research  Dresden, Helmholtzstr. 20, D-01069 Dresden, Germany}
\affiliation {Institut f\"ur Festk\"orperphysik,  Technische Universität Dresden, D-01062 Dresden, Germany}
\author{T.\,Wolf}
\affiliation{Institute for Quantum Materials and Technologies, Karlsruhe Institute of Technology, 76021 Karlsruhe, Germany}
\author{F.\,Hardy}
\affiliation{Institute for Quantum Materials and Technologies, Karlsruhe Institute of Technology, 76021 Karlsruhe, Germany}
\author{ C.\,Meingast}
\affiliation{Institute for Quantum Materials and Technologies, Karlsruhe Institute of Technology, 76021 Karlsruhe, Germany}
\author{H.S.\,Jeevan}
\affiliation{Institut f\"ur Physik, Universit\"at Augsburg, Universit\"atstr.1, D-86135 Augsburg, Germany}
\author{J.\,Maiwald}
\altaffiliation{Present address: Quantum Matter Institute, University of British Columbia, Vancouver, BC V6T 1Z4, Canada}
\affiliation{Institut f\"ur Physik, Universit\"at Augsburg, Universit\"atstr.1, D-86135 Augsburg, Germany}
\author{ P.\,Gegenwart}
\affiliation{Institut f\"ur Physik, Universit\"at Augsburg, Universit\"atstr.1, D-86135 Augsburg, Germany}
\author{ C.\,Felser}
\affiliation {Max Planck Institute for Chemical Physics of Solids, D-01187 Dresden, Germany}
\author{B.\,B\"uchner}
\affiliation{Leibniz Institute for Solid State and Materials Research  Dresden, Helmholtzstr. 20, D-01069 Dresden, Germany}
\affiliation {Institut f\"ur Festk\"orperphysik,  Technische Universität Dresden, D-01062 Dresden, Germany}

\date{\today}

\begin{abstract}\
We report an angle-resolved photoemission study of a series of hole and electron doped iron-based superconductors, their parent compound BaFe$_2$As$_2$, and their cousins BaCr$_2$As$_2$ and BaCo$_2$As$_2$. We focus on the inner hole
pocket, which is the hot spot in these compounds. More specifically, we determine the energy (E) dependent scattering rate $\Gamma (E)$ as a function of the $3d$ count. Moreover, for the compounds K$_{0.4}$Ba$_{0.6}$Fe$_2$As$_2$ and BaCr$_2$As$_2$ we derive the energy dependence  of the renormalization function $Z(E)$ and the imaginary part of the self-energy function $\Im\Sigma(E)$. We obtain a  non-Fermi-liquid-like linear in energy scattering rate $\Gamma(E\gg  k_{\mathrm{B}}T)$, independent of the dopant concentration. The main result is that the slope $\beta=\Gamma (E \gg k_{\mathrm{B}}T)/E$, reaches its  maxima near optimal doping and   scales with the superconducting transition temperature. This supports the spin fluctuation model for superconductivity for these materials. In the optimally hole-doped compound, the slope of the scattering rate of the inner hole pocket is about three times bigger than the Planckian limit $\Gamma (E)/E \approx 1$. This result together with the energy dependence of the renormalization function $Z(E)$ signals very incoherent charge carriers in the normal state which transform at low temperatures to a coherent unconventional superconducting state.

\end{abstract}


\maketitle

\section{\label{sec:intro} I. INTRODUCTION}

There is an ongoing debate about the mechanism for unconventional superconductivity in
strange metals. For the iron-based superconductors (FeSCs)\,\cite{Kamihara2008} the most popular model  is connected with spin fluctuation scattering processes between hole pockets in the center and electron pockets at the border  of the Brillouin zone (BZ), together with a sign change of the superconducting order parameter: the $s^{\pm}$ superconductivity~\cite{Mazin2008a,Kuroki2008}. Thus it is very interesting to study the energy ($E$) dependence of the scattering rates $\Gamma(E)$ on those sections of the Fermi surface which show the largest
superconducting order parameter (hot spots). Those studies as a function of doping concentration are the main topic of the present paper. In a local approximation, the scattering rates depend on  the effective  Coulomb interaction $U_{\mathrm{eff}}(U,J_{\mathrm{H}})$, where $U$ and $J_{\mathrm{H}}$ are the on-site Coulomb interaction and the Hund exchange interaction, respectively. Therefore, it is interesting to complement the scattering rates with results on correlation effects using other observables, e.g., the renormalization function $Z(E)$. Moreover, the investigation of correlation effects in multi-orbital systems as such is at present one of the great unresolved problems in solid state physics\,\cite{Haule2009,Aichhorn2009,Aichhorn2010,Medici2011,Werner2012,Razzoli2015,Bascones2016}. 

\par

The cuprates have just one band near the Fermi surface.  In the FeSCs there are three hole and two electron pockets, and several of them have sections with different orbital character [see Fig.~1(a)]. Therefore, it is rather difficult to receive a microscopic picture for  the mechanism of high-$T_{\mathrm{c}}$ superconductivity in these materials and one needs momentum dependent results which one can obtain from Angle-Resolved Photoemission Spectroscopy (ARPES)~\cite{Damascelli2003,Sobota2020}.

\par

\par
\begin{figure*}[t]
\centering
\includegraphics[angle=0,width=1.0\linewidth]{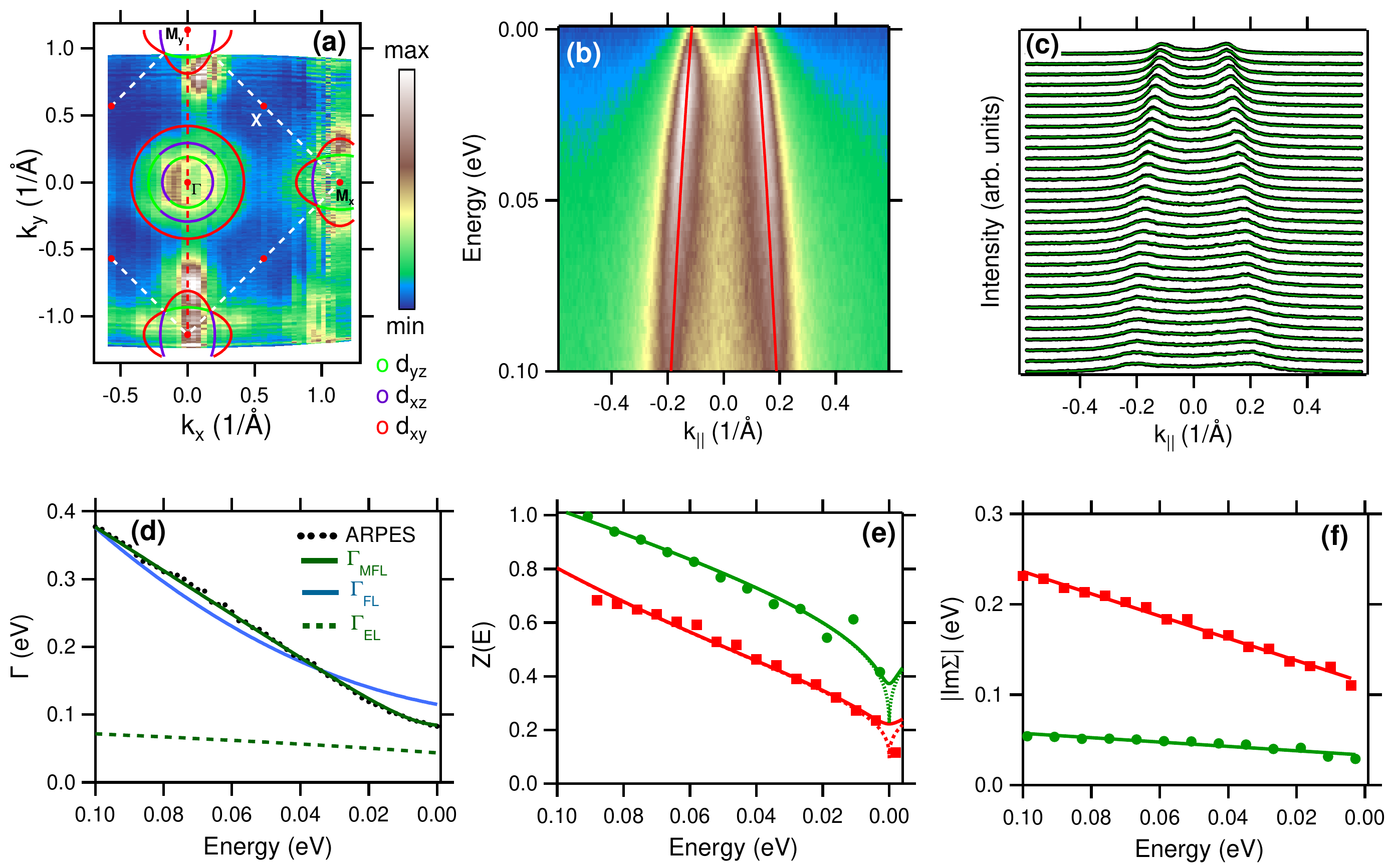}
 \caption{\label{edm}
ARPES data of optimally hole-doped K$_{0.4}$Ba$_{0.6}$Fe$_2$As$_2$ in the normal state at $T=50$ K measured with $h\nu=75$ eV vertically polarized photons. (a) Fermi surface derived from ARPES spectral weight near the Fermi level. The data are overlayed with a schematic plot of the Fermi surface of the three hole pockets near $\Gamma$ and two electron pockets near the M points. The predominant orbital character is indicated by colors.  (b) Experimental energy-momentum distribution map, measured along the red dashed line in (a). Red line: fitted dispersion. (c) Waterfall plot of the momentum distribution curves (black points) together with a  least squares fit (green lines). Uppermost spectrum $E=E_{\mathrm{F}}$, lowest spectrum  $E=0.12$ eV. (d) Experimental scattering rates $\Gamma(E,T=50$ K$)$ (black points) as a function of energy  derived from the least squares fit of the intensity distribution. Green line: fit with a theoretical non-Fermi-liquid  model (see text). Dashed green line: estimated contribution from elastic scattering. Blue line: fit with a Fermi liquid behavior. (e) Renormalization function Z(E) derived from ARPES data (red squares). Solid red line: fit to a function given in Eq. (3). The results are compared with analogous data from BaCr$_2$As$_2$ (green points,  solid green line). The dotted lines are extrapolations of the fit curves to zero temperature. (f) Imaginary part of the self energy $\Im\Sigma(E)$ derived from our experimental values  $\Gamma(E)$ and $Z(E)$ using the relation  $\Gamma(E)=-2Z(E)\Im\Sigma(E)$ (red squares). Solid red line: fit to a linear energy dependence. The results are compared with analogous data from BaCr$_2$As$_2$ (green points, solid green line).
}
\centering
\end{figure*}

\par
 
In the present contribution, we focus on the doping dependence  of the strength of the scattering rates $\Gamma$  of the inner hole pocket. This pocket corresponds to a hot spot exhibiting the highest isotropic superconducting gap~\cite{Ding2008} and  according to our previous studies on various FeSCs shows the highest scattering rates when compared with the other pockets~\cite{Fink2015,Fink2017,Avigo2017,Fink2019}. We obtain a doping dependence with a pronounced \textit{maximum} near optimal hole doping for both, the  superconducting transition temperature $T_{\mathrm{c}}$ and for the slope $\beta=\frac{\Gamma(E \gg k_{\mathrm{B}}T)}{E}$. We therefore conclude that the scattering rates of the inner hole pockets are  related to superconductivity. In this way we support the $s^\pm$ model~\cite{Mazin2008a,Kuroki2008}.  From this comparison we also derive that superconductivity is determined by a combination of correlation effects and nesting conditions.
Furthermore, using our new elaborated evaluation technique for the analysis of $\Gamma(E)$ and the renormalization function $Z(E)$, we conclude that the charge carriers at the hot spots have non-Fermi-liquid character. Finally, from our experimental scattering rates of optimally hole doped FeSCs we conclude that in particular for the optimally hole doped compound, the hot spot charge carriers  are heavily incoherent, i.e., $\beta \approx3$ well above the Planckian  limit of $\beta=1$~\cite{Zaanen2004}.

\par

\section{\label{sec:exp} II. EXPERIMENTAL}

 Single crystals were grown  using the self-flux technique\,\cite{Morozov2010,Hardy2016,Maiwald2012}.
ARPES measurements were conducted at the $1^2$ and $1^3$-ARPES end stations attached to the beamline UE112 PGM at BESSY. All data presented in this contribution were taken in the normal state at temperatures between 5 and 50 K.   The achieved energy and angle resolutions were between 4 and 15 meV and 0.2$^\circ$, respectively. Polarized photons were used and an attempt was made to achieve prefect alignment of the  sample  to select spectral weight with a specific orbital character and to avoid a contamination from other bands\,\cite{Fink2009,Moser2017}. Photon energies were varied between $h\nu=20-130$ eV  to reach  $k_z$ values close to $\Gamma$ points in the BZ.   An inner potentials  of 14 eV was used to calculate the $k_z$ values from the photon energy. 

\section{\label{sec:dat} III: DATA ANALYSIS}

Using ARPES one measures the product of the spectral function~\cite{Mahan2000},   with a transition matrix element, and the Fermi function. This product is convoluted with the energy and momentum resolution\,\cite{Damascelli2003,Sobota2020}. The spectral function  is given by 
\begin{equation}
A(E,\mathbf{k},T)=\frac{1}{\pi}Z(E,\mathbf{k},T)\frac{\frac{\Gamma(E,\mathbf{k},T)}{2}}
{[E -\epsilon _{\mathbf{k}}^*]^2+[\frac{\Gamma(E,\mathbf{k},T)}{2}]^2}.
\label{26_1}
\end{equation}
The spectral function $A$, the scattering rate $\Gamma$, and the renormalization function $Z$ are energy, momentum ($\mathbf{k}$), and temperature ($T$) dependent.
$\Gamma$ is related to the complex self-energy $\Sigma$ by the relation  $\Gamma=-2Z\Im\Sigma$. Z is connected to the $\Re\Sigma$ by  
$Z=1/(1+\frac{\Re\Sigma}{E})$. $\epsilon_{\mathbf{k}}$ and $\epsilon_{\mathbf{k}}^{\star}=Z\epsilon_{\mathbf{k}}$ are  the bare particle dispersion and the renormalized dispersion, respectively~\cite{Mahan2000,Damascelli2003,Sobota2020}. 
For a Fermi liquid,  $\Gamma_{\mathrm{FL}} \ll E$,  $Z(E)=\frac{1}{1+\lambda_{\mathrm{FL}}}=\frac{m}{m^*}=$const with $\frac{m^*}{m}$ equal to the mass enhancement.  $\Gamma_{\mathrm{FL}}$ which  is related to the inverse life time of the quasi-particles, is quadratic in energy and  in temperature.

\par
 
Very often the description of the spectral function is extended to higher scattering rates, i.e., $\Gamma$ is comparable to the binding energy $E$~\cite{Grimvall1981,Valla1999}. For this case, there is the conjecture of a marginal Fermi liquid~\cite{Varma1989}. There the complex self-energy is given by 
\begin{equation}
\Sigma_{\mathrm{MF}}(E)=\lambda_{\mathrm{MF}}(E\ln(|\frac{E_{\mathrm{c}}}{x}|)+i\frac{\pi}{2}x).
\end{equation} 
$x=\mathrm{max}(|E|,k_{\mathrm{B}}T)\approx (E^2+(\frac{\pi}{2}k_{\mathrm{B}}T)^2)^\frac{1}{2}$ and $E_{\mathrm{c}}$ is a cut-off energy.
In this case the renormalization function is energy dependent~\cite{Varma2020}: 
\begin{equation}
Z_{\mathrm{MF}}(E)=\frac{1}{1+\lambda_{\mathrm{MF}}\ln(|\frac{E_{\mathrm{c}}}{x}|)}.
\end{equation} 
We emphasize that for $E\gg k_{\mathrm{B}}T$, $\Im\Sigma_{MF}$ is linear in energy.

In the standard procedure for the evaluation of the scattering rates from the ARPES data, the momentum distribution curves (MDC, cuts at constant energy)  are fitted by a Lorentzian and the FWHM momentum width is multiplied by the velocity $\mathrm{d}\epsilon^*_k/\mathrm{d}k$ yielding $\Gamma(E,T)$~\cite{Valla1999,Valla1999b,LaShell2000,Bogdanov2000,Kaminski2001,Damascelli2003,Kordyuk2006a,Brouet2016,Cao2016,Reber2019}. Very often non-Lorentzian MDCs or energy distribution curves (EDC, cuts at constant k) are realized in the ARPES data. For example a  strong increase of $\Gamma(E,T)$ at high energies leads to long tails in the EDCs.   Furthermore, large scattering rates lead to contributions from the unoccupied part of the band causing  an apparent back-dispersion for $k<k_\mathrm{F}$ in hole pockets~\cite{Huefner1999} similar to dispersions in the superconducting state [see Fig.~1 (b)]. All this may lead to incorrect scattering rates when using the standard evaluation.

\par

To derive more exact data for the scattering rates, we developed a new elaborate evaluation method. We fit the two-dimensional $E-k$ intensity distribution map at once, using as parameters $\Gamma(E_n,T)$ at each energy point $E_n$ as well as parameters describing the dispersion of $\epsilon_k^*$. The assumption of a  parabolic dispersion gives reasonably fits. Higher orders in $k$ did not improve them. Other fit parameters
describe a weakly momentum and energy dependent background which is added to the spectral function. The sum is multiplied with the Fermi function and the product is then convoluted with the energy and momentum resolution. We mention that very close to the Fermi level, the derived $\Gamma$ values are very sensitive to the exact position of the Fermi level.  Normally, for the renormalization function we use two parameters  describing 
a polynominal energy dependence. In two cases, i.e., Ba$_{0.6}$K$_{0.4}$Fe$_2$As$_2$ and BaCr$_2$As$_2$, we use for each energy the parameters $Z(E_n)$ [see Fig.~1(e)]. The error bars for the scattering rates are not statistical errors, but the errors are estimated from the deviations from a mean value of several measurements using different spectrometers and photon energies.

\par

From the analysis of the ARPES data described above,  we obtain the energy dependent total scattering rate  $\Gamma(E,T)$ which is the sum of the elastic scattering rate  $\Gamma_{\mathrm{el}}(E)$  and the inelastic scattering rate $\Gamma_{\mathrm{in}}(E,T)$. To obtain the inelastic scattering rates we have to subtract the elastic scattering rates.
$\Gamma_{\mathrm{el}}(E)$ is usually assumed to be constant~\cite{Valla1999} which is reasonably in a semi-classical picture.
For a linear dispersion, when the velocity $v(E)$ is constant,   $\Gamma_{\mathrm{el}}(E)= v(E)w_0$ is constant. The momentum width at the Fermi level $w_0$ is related to the constant inverse mean free path between scattering sites inducing elastic scattering. In the case of  a parabolic dispersion, one has to subtract the energy dependent $\Gamma_{\mathrm{el}}(E)$ from 
$\Gamma(E,T)$.  In a semi-classical picture, $w_0$  can be calculated from the data near $E_{\mathrm{F}}$ because at small temperatures,  $\Gamma_{\mathrm{in}}(0,0)$ is zero and thus $\Gamma_{\mathrm{el}}(0)=\Gamma(0,0)= v(0)w_0$. Thus we derive for the energy dependence of the elastic scattering
$\Gamma_{\mathrm{el}}(E)= v(E)w_0=\Gamma(0,0)(v(E)/v(0))$. The velocities are taken from the measured dispersion.
Because it is not  clear whether the semi-classical model for the elastic scattering is really applicable for a highly correlated electron system,
in this contribution, we prefer to show  the full scattering rate $\Gamma(E,T)$.

\section{\label{sec:res} V. RESULTS}

In Fig.~\ref{edm}  we exemplary show  ARPES data of optimally hole-doped K$_{0.4}$Ba$_{0.6}$Fe$_2$As$_2$  ($T_{\mathrm{c}}=38$ K), recorded at a temperature $T=50$ K. Fig.~\ref{edm}(a) shows the measured Fermi surface, derived from the intensity close to the Fermi level. The panel also presents a schematic plot of the Fermi surface in which the predominant Fe $3d$ orbital character of the sections are indicated  in colors. In Fig.~\ref{edm}(b) we show an energy-momentum distribution map of the measured intensity measured along the high symmetry direction $\Gamma$-M$_y$  [see the red dashed line in Fig.~\ref{edm}(a))]. Under these experimental conditions, near the $\Gamma$ point we record  the spectral weight of the inner hole pocket having predominantly Fe $3d_{yz}$ character~\cite{Fink2009}. 

\par

 A waterfall plot of MDCs together with fits is shown in Fig.~1(c) for the energy range $0 \leq E \leq 0.12$ eV.  From those fits, we obtain the dispersion $\epsilon^*_k$, the scattering rates $\Gamma(E,T)$, and the renormalization function $Z(E)$.
We emphasize that our evaluation method for $\Gamma(E,T)$ is much better suited for data close to the top of bands or close to the Fermi level than the standard evaluation method.

\par

The derived dispersion $\epsilon^*_k$ is shown as a red line in Fig.~\ref{edm}(b). In Fig.~\ref{edm}(d) we present the derived experimental $\Gamma(E)$ together with fits. The blue line is related to a fit for a Fermi liquid [$\Gamma(E)=\Gamma(E)_{\mathrm{in}}+\Gamma(E)_{\mathrm{el}}$ with $\Gamma(E)_{\mathrm{in}}=\beta_{\mathrm{FL}}(E^2+(\frac{\pi}{2}k_{\mathrm{B}}T)^2)$]. The green line corresponds to   a non-Fermi liquid fit with $\Gamma(E)_{\mathrm{in}}=\beta x$ [see Eq. (2)]. $\Gamma(E)_{\mathrm{el}}$ is  due to impurities or defects at the surface~\cite{Damascelli2003,Sobota2020}. Using a parabolic band and a constant mean free path we derive a slightly energy dependent $\Gamma(E)_{\mathrm{el}}$ [see the  dashed green line in  Fig.~\ref{edm}(d)]. 

\par

The Fermi liquid fit does not describe the experimental results. On the other hand, the non-Fermi liquid fit with $\beta=3.3$ is much closer to the ARPES data.  $\Gamma(0,T=50~$K$)=0.084$ eV  and  $\Gamma_{\mathrm{el}} (0)=0.044$~eV.  The difference $\Gamma(0,T=50~$K$) -\Gamma_{\mathrm{el}}(0)=0.040$~eV agrees well with the contribution $\Gamma_{\mathrm{in}}(0,T=50~$K$)$ corresponding to a finite temperature $T=50$ K equal to $\beta \frac{\pi}{2} k_{\mathrm{B}}T=0.042$~eV. $\Gamma_{\mathrm{el}} (0)=0.044$~eV together with the Fermi velocity $v_{\mathrm{F}}=0.9$~eV\AA,  corresponds to a mean free path of $\approx 60~$\AA
, a value which is quite common in ARPES experiments, also for not strongly correlated materials ~\cite{Nicolay2006,Valla1999}.
 
\par

In Fig.~\ref{edm}(e) we depict the experimental renormalization function $Z(E,T)$  together with a fit using Eq. (3). In that panel  we compare the data of  K$_{0.4}$Ba$_{0.6}$Fe$_2$As$_2$  with those of BaCr$_2$As$_2$, which is a less correlated material (see below). For K$_{0.4}$Ba$_{0.6}$Fe$_2$As$_2$ and BaCr$_2$As$_2$
we derive $\lambda_{\mathrm{MF}}=1.0$ and
0.4 and cutoff energies $E_{\mathrm{c}}$ equal to  0.13~eV and 0.1~eV, respectively.

\par

In the following we show ARPES data for various electron and hole doped FeSCs, the undoped antiferromagnetic parent compound BaFe$_2$As$_2$, and the cousin compounds  BaCr$_2$As$_2$ and BaCo$_2$As$_2$. We depict the energy-momentum distribution maps together with  fit results of the dispersion (solid red line) measured along the high symmetry direction $\Gamma$-M$_y$,  waterfall plots of the MDCs  together with   fit results, and energy dependent scattering rates $\Gamma(E\gg  k_{\mathrm{B}}T)$, together with  a fit using a linear energy dependence of $\Gamma$. The energy range of the waterfall plots is listed after the chemical formula of the compounds. 

\begin{figure}[t]
\centering
\includegraphics[angle=0,width=1.0\linewidth]{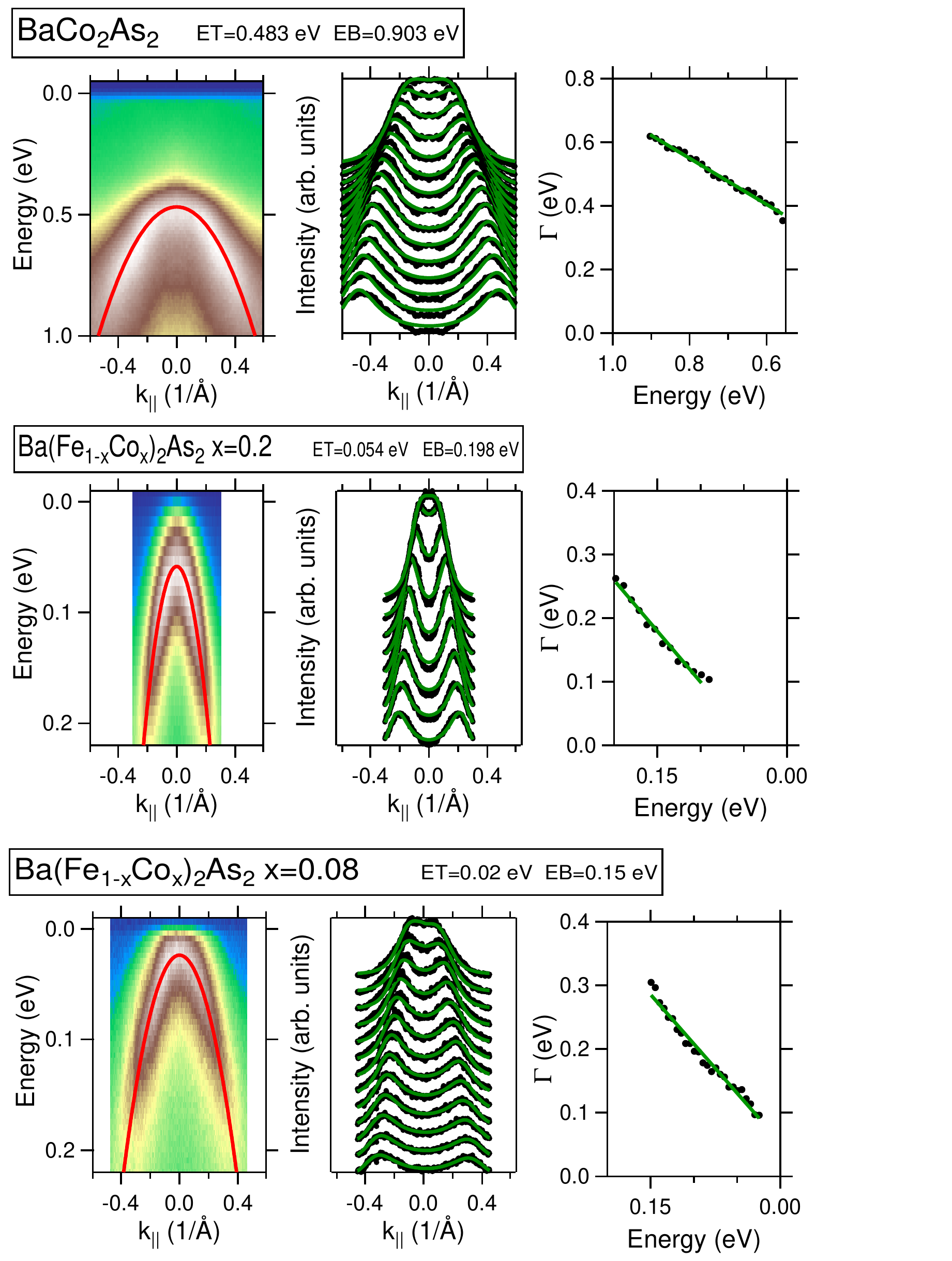}
 \caption{\label{sup1}Upper row: data for BaFe$_2$As$_2$. Left panel: ARPES energy-momentum map together with a fit of the dispersion (red line). Middle panel: waterfall plots of the ARPES MDCs (black dots) together with least squares fit results (green line). Right panel $\Gamma(E\gg  k_{\mathrm{B}}T,T)$ derived from ARPES data (black dots) together with fit results (green line).  Middle row: analogous data for  Ba(Fe$_{1-x}$Co$_{x})_2$As$_2$ $x=0.2$. Lower row: analogous data for Ba(Fe$_{1-x}$Co$_{x})_2$As$_2$ $x=0.08$.
}
\centering
\end{figure}

In Fig.~\ref{sup1} we depict  data  of BaCo$_2$As$_2$ (upper row) measured with a photon energy h$\nu=127$ eV and a temperature of $T=54$ K. The middle row shows analogous results for  Ba(Fe$_{1-x}$Co$_{x})_2$As$_2$, $x=0.2$ (h$\nu=49$ eV, $T=50$ K). The lower row shows analogous results for  Ba(Fe$_{1-x}$Co$_{x})_2$As$_2$ $x=0.08$ (h$\nu=49$ eV, $T=30$ K).

\begin{figure}[t]
\centering
\includegraphics[angle=0,width=1.0\linewidth]{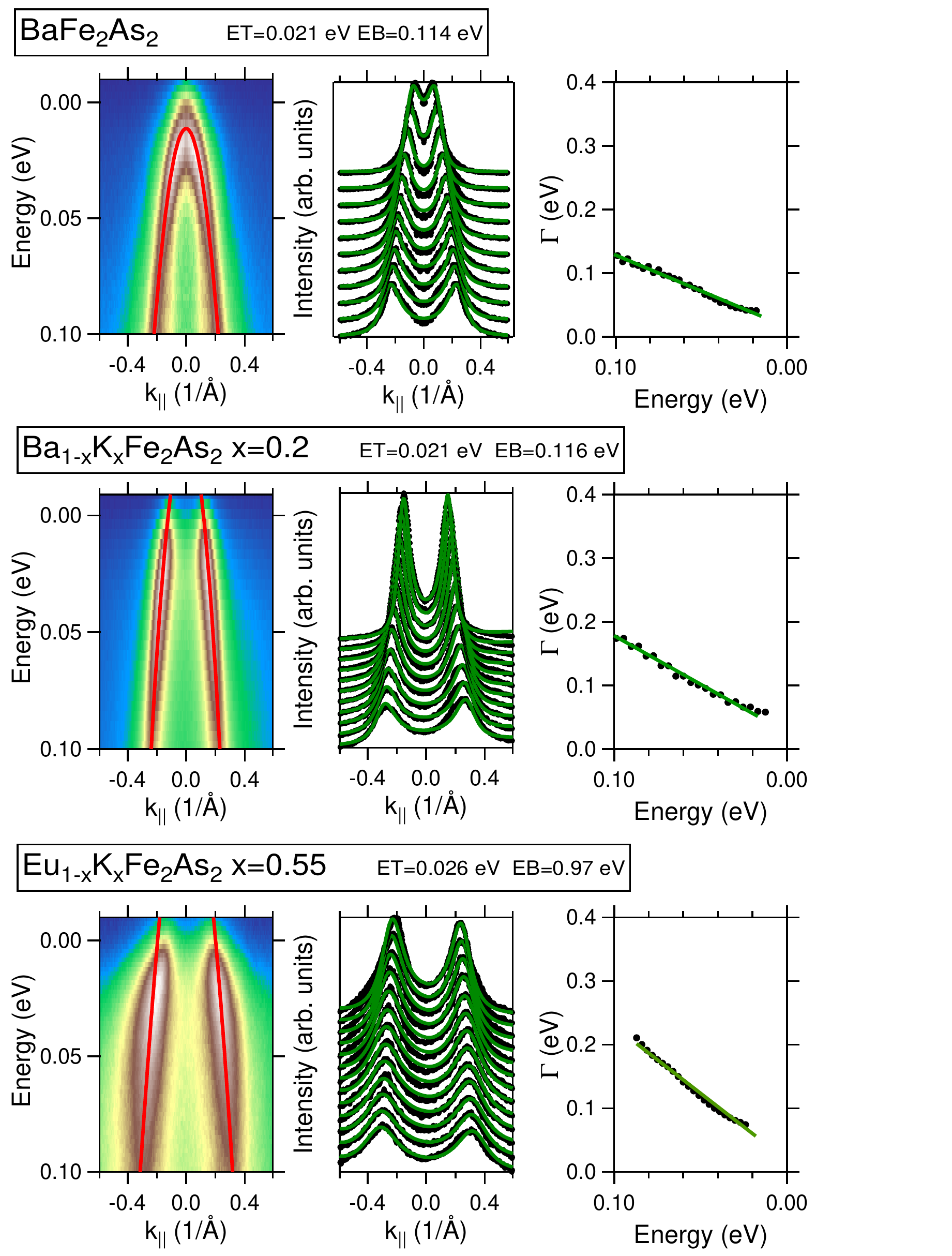}
 \caption{\label{sup2}Upper row: data for BaFe$_2$As$_2$. Left panel: ARPES energy-momentum map together with a fit of the dispersion (red line). Middle panel: waterfall plots of the ARPES MDCs (black dots) together with least squares fit results (green line). Right panel $\Gamma(E\gg  k_{\mathrm{B}}T,T)$ derived from ARPES data (black dots) together with fit results (green line). Middle row: analogous data for Ba$_{1-x}$K$_x$Fe$_2$As$_2$ $x=0.2$. Lower row: analogous data for Eu$_{1-x}$K$_x$Fe$_2$As$_2$ $x=0.55$.
}
\centering
\end{figure}

In Fig.~\ref{sup2} we show  data  of BaFe$_2$As$_2$ (upper row) measured with a photon energy h$\nu=78$ eV and a temperature of $T=56$ K. The middle row shows  results for  Ba$_{1-x}$K$_x$Fe$_2$As$_2$, $x=0.2$ (h$\nu=78$ eV, $T=55$ K). The lower row shows  results for  Eu$_{1-x}$K$_x$Fe$_2$As$_2$, $x=0.55$ (h$\nu=40$ eV, $T=40$ K).

\begin{figure}[t]
\centering
\includegraphics[angle=0,width=1.0\linewidth]{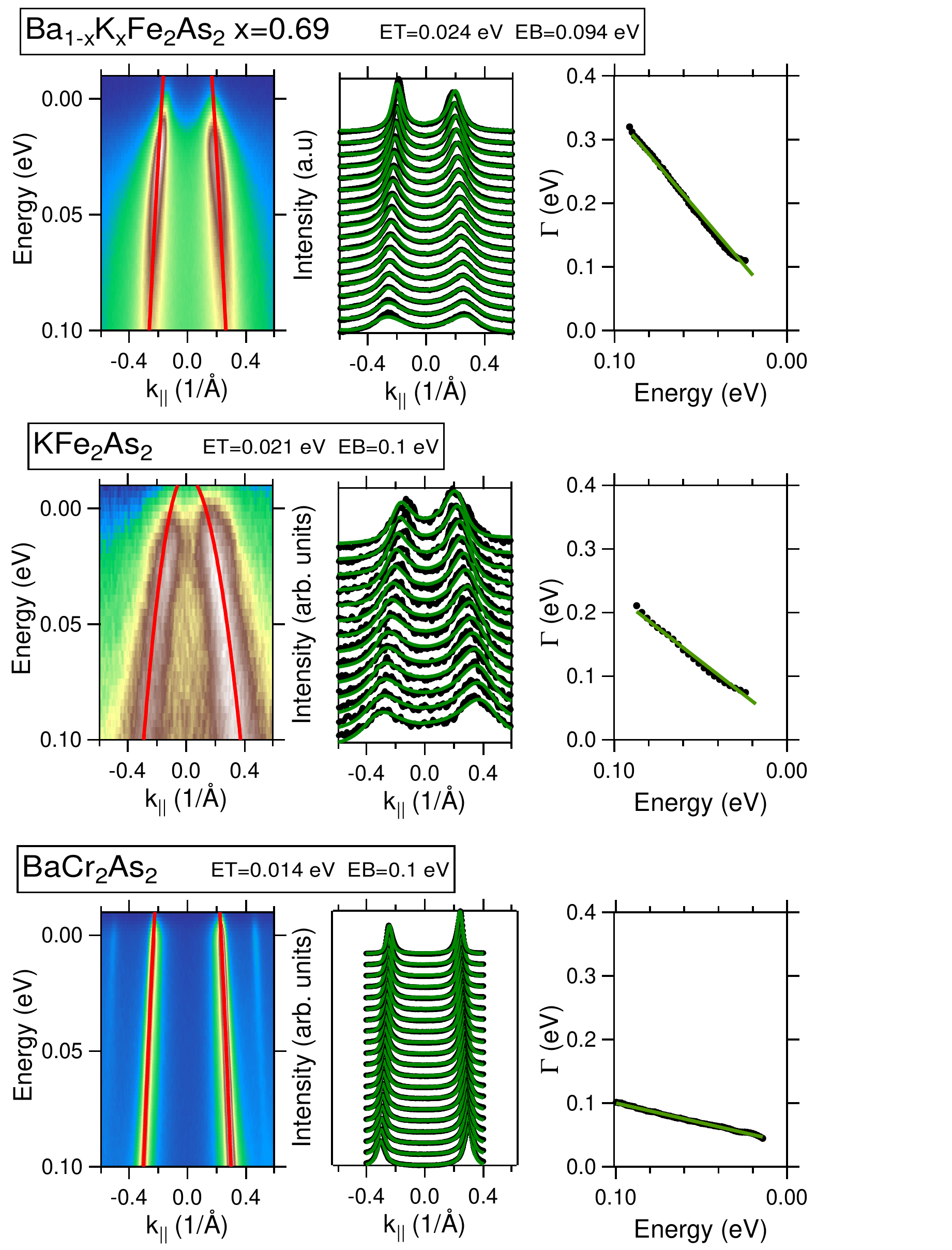}
\caption{\label{sup3}Upper row: data for Ba$_{1-x}$K$_x$Fe$_2$As$_2$ $x=0.69$. Left panel: ARPES energy-momentum map together with a fit of the dispersion (red line). Middle panel: waterfall plots of the ARPES MDCs (black dots) together with least squares fit results (green line). Right panel $\Gamma(E\gg  k_{\mathrm{B}}T,T)$ derived from ARPES data (black dots) together with fit results (green line). Middle row: analogous data for KFe$_2$As$_2$. Lower row: analogous data for BaCr$_2$As$_2$.
}
\centering
\end{figure}

In Fig.~\ref{sup3} we show  in the upper row data together with fit results of Ba$_{1-x}$K$_x$Fe$_2$As$_2$ $x=0.69$ (h$\nu=71$ eV, $T=55$ K). The middle row shows  results for  KFe$_2$As$_2$ (h$\nu=68$ eV, $T=22$ K) and the lower row depicts data for BaCr$_2$As$_2$ (h$\nu=70$ eV, $T=50$ K).
In the latter, a low intensity dispersion is also visible from the middle hole pocket.

\begin{figure}[t]
\centering
\includegraphics[angle=0,width=1.0\linewidth]{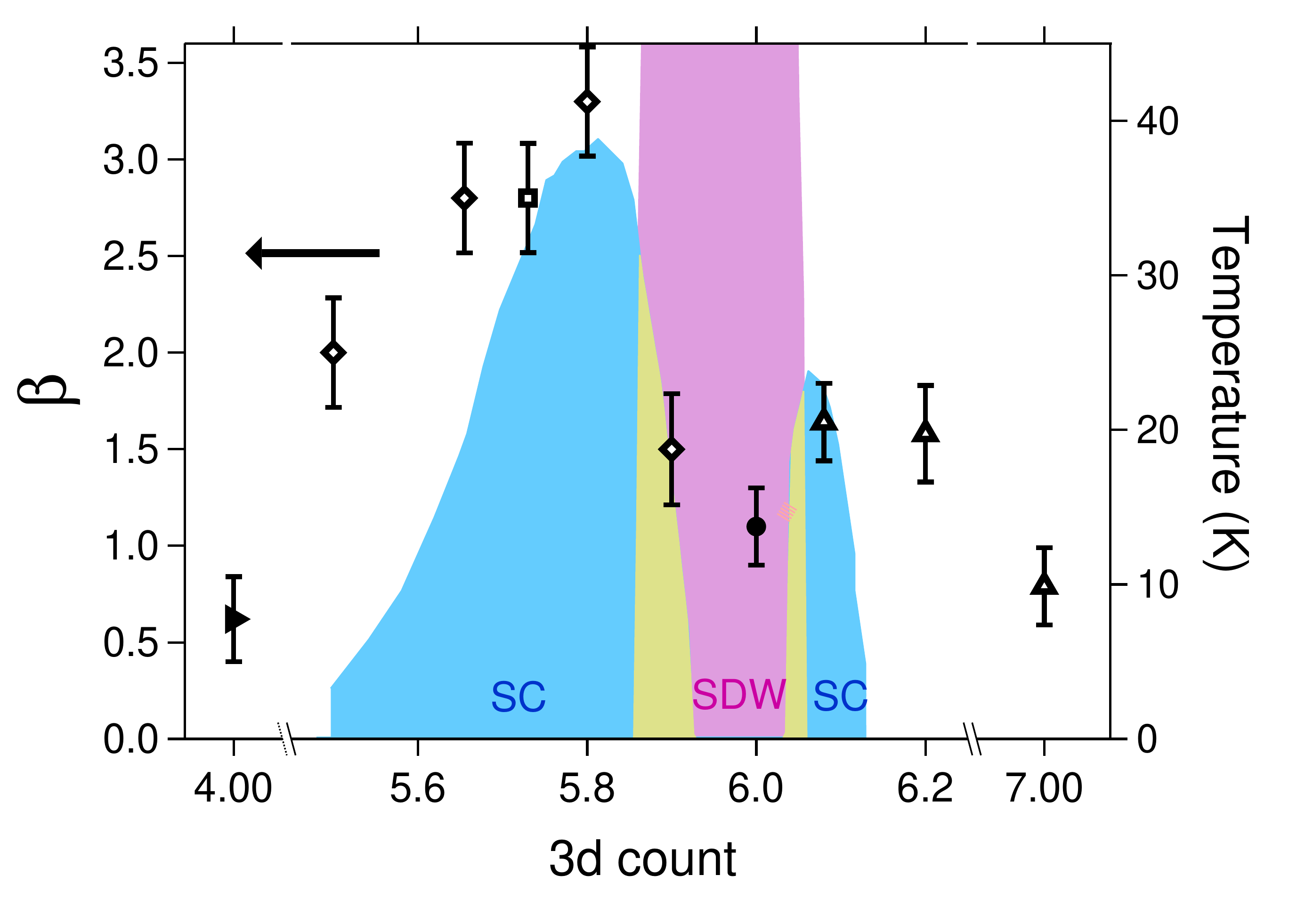}
\caption{Experimental values of the slope of the scattering rate $\beta=\Gamma(E\gg  k_{\mathrm{B}}T,T)/E$  as a function of $3d$ count of various FeSCs and related compounds. Part of the data have been presented already in Refs~\cite{Fink2015,Fink2017,Nayak2017,Fink2019}.
$\diamond$~Ba$_{1-x}$K$_x$Fe$_2$As$_2$, $\square$~Eu$_{1-x}$K$_x$Fe$_2$As$_2$,
$\bigtriangleup~$ Ba(Fe$_{1-x}$Co$_x$)$_2$As$_2$,
$\blacktriangleright$~BaCr$_2$As$_2$, $\bullet$~BaFe$_2$As$_2$. The filled symbols correspond to antiferromagnetic compounds.
The ARPES data are underlaid with a phase diagram for hole and electron doped BaFe$_2$As$_2$ compounds with the range of the superconducting phase (SC, blue), the spin density wave phase (SDW, magenta), and the range where both phases overlap (yellow)~\cite{Hardy2016}.
}
\label{Gam_3d}
\centering
\end{figure}

 The results of $ \beta $ together with a phase diagram~\cite{Hardy2016} as a function of the $3d$ count, i.e., the number of $3d$ electrons derived from the chemical composition, are presented in Fig.~\ref{Gam_3d}.

\par

\section{\label{sec:dis} IV. DISCUSSION}
Regions with a linear increase of the scattering rates as a function of energy or temperature have been  detected in resistivity\,\cite{Hussey2008,Taillefer2020,Zaanen2019} and  ARPES data\,\cite{Valla1999b,Bogdanov2000,Kaminski2001,Koitzsch2004,Kordyuk2006,Reber2019,Fink2015,Fink2017,Sanchez-Barriga2018,Fink2019,Avigo2017,Nayak2017} in various correlated metals. 
In this contribution  we emphasize that for the hot spot of Ba$_{1-x}$K$_x$Fe$_2$As$_2$, both from the energy and the influence of a finite temperature, an incoherent non-Fermi-liquid behavior is observed.  The linear energy dependence of $\Gamma(E\gg  k_{\mathrm{B}}T)$ in the full range of the $3d$ count from 4 to 7 is different from that at the nodal point in cuprates, where for the scattering rate a continuous superposition of a linear and a quadratic energy dependence~\cite{Hussey2008} or a $T^n$ dependence with $n$ changing from one to two was discussed~\cite{Koitzsch2004}, when moving from optimal  to overdoped compounds.   

\par

It is interesting to note that the largest slope of the scattering rate occurs near the optimally hole-doped compound. For this $3d$ count the scattering rate is about three times bigger than the energy and therefore well above the Planckian limit where $\Gamma_{\mathrm{in}}(E,T\approx 0)=E$~\cite{Zaanen2004,Zaanen2019}. Moreover, it is also remarkable that the slope, which to our knowledge is the largest slope ever detected by ARPES, is about three times bigger than that in optimally doped cuprate superconductors along the diagonal direction~\cite{Valla1999b,Kaminski2001,Yusof2002,Koitzsch2004,Reber2019}. It is certainly a challenge to understand how the completely incoherent hot spots in the normal state  transform into a coherent superconducting state.

\par

Looking to Fig.~\ref{Gam_3d}, we detect a clear linkage of the scattering rates with the superconducting transition temperatures $T_{\mathrm{c}}$. There is a clear \textit{maximum}  near optimally hole doping. This strongly supports the scenario that superconductivity is induced by spin flip interband transitions between hole pockets and electron pockets~\cite{Mazin2008a,Kuroki2008}. The reduced scattering rates for BaCr$_2$As$_2$ and BaFe$_2$As$_2$ can be probably explained by the existence of a magnetic order. 

Using Fermi's Golden rule and a local approximation, the scattering rate is related to the on-site interaction $U_{\mathrm{eff}}$ and the charge susceptibility which determines the relaxation of the photoelectron to lower energies by an electron-hole excitations~\cite{Echenique2000,Monney2012}.
According to Avigo \textit{et al.}~\cite{Avigo2017}, the top of the inner hole  pocket and the bottom of the inner electron pocket are separated in these compounds by $\approx 0.1$ eV. In a rigid band approximation, the Fermi level moves $\approx$ 0.5 eV per dopant electron/hole.
Assuming that  intra-band transitions would only be  possible when hole and electron pockets cross the Fermi level, these transitions  could only occur in a range of a $3d$ count of $\pm$ 0.2  around the undoped sample. This is in line with the maximum of $T_{\mathrm{c}}$ for the hole doped system but too large for the electron doped systems.

\par

 Superconductivity at higher dopant concentration could be explained by the fact that the equation for the superconducting transition temperature yields also solutions for excitations away from the Fermi level in a range of the coupling energies. If these excitations are spin fluctuations, which have  an energy range between 0.01 to 0.2 eV~\cite{Johnston2010,Lumsden2010}, we can understand that  there is also a finite $T_{\mathrm{c}}$ in the over-doped compounds. On the other hand, the decrease of $T_{\mathrm{c}}$ can be understood by a decrease of the spin susceptibility, also detected by inelastic neutron scattering~\cite{Lumsden2010}, because in the overdoped compounds, the electron or hole  pocket has moved far  above or below the Fermi level, respectively. This reduction of the scattering rate  is clearly detected for  strongly over-doped compounds. On the other hand, thermal properties have derived a strong enhancement of the effective mass when going to higher hole doping~\cite{Hardy2016}. This indicates that the related flat bands close to the Fermi level, detected  in KFe$_2$As$_2$ by ARPES~\cite{Yoshida2014} and by quantum-oscillation experiments~\cite{Terashima2009a}, are causing a large band mass enhancement, but not superconductivity.

\par
  
For a $3d$ count of four (BaCr$_2$As$_2$ ~\cite{Nayak2017,Richard2017}, see Fig.~\ref{Gam_3d}) as well as for seven or eight (BaCo$_2$As$_2$ ~\cite{Xu2013b} or BaNi$_2$As$_2$ ~\cite{Pavlov2019}), reduced scattering rates  or correlation effects are detected.
A  reduction  of correlation effects at higher electron doping  was also derived from ARPES results on the mass enhancement in BaCo$_2$As$_2$ ~\cite{Xu2013b} and BaNi$_2$As$_2$ ~\cite{Pavlov2019}.
 
As explained in Ref.~\cite{Fink2017} the larger scattering rate for the hole doped systems, when compared to the electron systems, can be explained by an enhancement of $U_{\mathrm{eff}}$ in the proximity to the half-filled $3d$ shell due to Hund exchange coupling ~\cite{Haule2008,Medici2015}. This means that our experimental data indicate that the on-site Coulomb interaction $U$, the Hund exchange interaction $J_H$, the band filling, and the nesting conditions determines the strength of the correlation effects in the inner hole pocket, which shows the highest superconducting gap.

As shown in our previous work on electron doped Ba(Fe$_{1-x}$Co$_x$)$_2$As$_2$ and BaFe$_{1-x}$(Co,Rh)$_x$As$_2$~\cite{Fink2015,Fink2017}, on undoped LiFeAs~\cite{Fink2019}, and on hole doped  Ba$_{1-x}$K$_x$Fe$_2$As$_2$, $x=0.4$~\cite{Fink2017}, the  middle hole pocket ($xz/yz$ character) show scattering rates which are reduced by a factor of two to three when compared with the inner hole pocket ($yz/xz$ orbital character). This tendency is also observed in some hole doped Ba$_{1-x}$K$_x$Fe$_2$As$_2$ compounds for $x$ different from 0.4 (not shown). Moreover, also in our recent work on EuRbFe$_4$As$_4$ this difference in the scattering rate between the inner and the middle hole pockets was detected~\cite{Hemmida2020}. Because the spectral weight of the outer hole pocket ($xy$ character) is very sensitive to impurities or dopant atoms, we have data for this pocket only for BaFe$_{1-x}$(Co,Rh)$_x$As$_2$~\cite{Fink2015,Fink2017} and LiFeAs~\cite{Fink2019}. There, we also detect a reduced scattering rate of the outer hole pocket when compared with the inner hole pocket. We point out that on the middle and the outer hole pockets, also smaller superconducting gaps have been detected~\cite{Ding2008}. This supports our suggestion, that the scattering rates are related to the superconducting order parameter.
 
With increasing electron doping by Co,
the stronger reduction of $T_c$ compared to the weaker decay of the scattering rates  for overdoped Ba(Fe$_{1-x}$Co$_x$)$_2$As$_2$ can be explained by a large pair-breaking by Co scatterers~\cite{Hardy2016}.
The observation of a finite scattering rate at high dopant concentrations can be explained by a scattering from the inner hole pocket to other bands close to the Fermi level.

\par

The renormalization function $Z(E)$ shows a strong energy and $3d$ count  dependence [see Fig.~\ref{edm}(e)] as expected for a non-Fermi liquid~\cite{Varma1989}.  The less correlated BaCr$_2$As$_2$ compound
(smaller $\beta$) shows a smaller $\lambda_{\mathrm{MF}}$ compared to the more correlated  K$_{0.4}$Ba$_{0.6}$Fe$_2$As$_2$. The cut-off energy $E_c$ for the latter compound is close to the extension of the Drude peak detected by optical spectroscopy~\cite{Yang2009d}. The observation of an almost equal cut-off energy for BaCr$_2$As$_2$ is not expected when one assumes that $E_{\mathrm{c}}$ is related to the band renormalization. The finite value of $Z(0)$ of K$_{0.4}$Ba$_{0.6}$Fe$_2$As$_2$ may be caused by the finite temperature.

\par

We have also calculated $\Im\Sigma(E)$  from the relation $\Gamma(E)=-2Z(E)\Im\Sigma(E)$ using our experimental values for $\Gamma(E)$ and $Z(E)$. The results for K$_{0.4}$Ba$_{0.6}$Fe$_2$As$_2$ and BaCr$_2$As$_2$ are presented in Fig.~1(f). For both compounds we obtain a linear energy dependence. The coupling constants derived from the slopes of $\Im\Sigma(E)$ are $\lambda_{\mathrm{MF}} = 0.8$ and 0.2 for K$_{0.4}$Ba$_{0.6}$Fe$_2$As$_2$ and BaCr$_2$As$_2$, respectively. It is interesting to compare the former result with semi-phenomenological  calculations on the basis of a coupling of the electrons to spin fluctuations, obtained for the inner hole pocket of K$_{0.4}$Ba$_{0.6}$Fe$_2$As$_2$ in the normal state~\cite{Heimes2011}: at higher energies  $\Im\Sigma(E) \propto E$ and the slope is close to one in good agreement with our experimental results.   We conclude that  our ARPES results on the energy  dependence and the finite temperature influence
of $\Gamma(E,T)$ and $Z(E)$ consistently support a charge carrier behavior close to a marginal Fermi liquid.

\section{ ACKNOWLEDGMENTS}
J.F. thanks  Luis Craco for helpful discussions. This work has been supported by the Deutsche Forschungsgemeinschaft (DFG) through the Priority Program SPP1458, through the Emmy Noether Programm in project WU595/3-3 (S.W.), through Research Training Group GRK 1621, and via Grant No. DFG AS 523/4-1 (S.A.).

\bibliographystyle{apsrev4-2}
\bibliography{Pnictide}

\end{document}